\documentclass[conference]{IEEEtran}
\IEEEoverridecommandlockouts
\usepackage[table]{xcolor}
\usepackage{color}
\usepackage{comment}
\usepackage{ragged2e}
\usepackage[Symbol]{upgreek}

\usepackage{soul}
\newcommand{\drop}[1]{\textcolor{red}{#1}}
\renewcommand{\drop}[1]{}
\definecolor{cadmiumgreen}{rgb}{0.0, 0.42, 0.24}
\usepackage{multirow}
\usepackage{enumitem}
\usepackage{fancyhdr}

\usepackage{graphicx}
\usepackage{booktabs} 
\usepackage{amsmath}
\usepackage{tikz}

\usepackage{circledsteps}
\usepackage{nicematrix}
\usepackage{pifont}
\usepackage{mathtools}

\usepackage[ruled, vlined, norelsize]{algorithm2e}
\SetKwInput{KwInput}{Input}
\SetKwInput{KwOutput}{Output}
\SetKwInput{KwInit}{Initialization}
\SetKwInput{Kwprocedure}{Procedure}
\usepackage{lipsum}
\usepackage{dblfloatfix}

\usepackage{bm}

\usepackage{scalerel}[2016/12/29]

\newcommand{\ssup}[2]{{#1}^{\scaleobj{0.8}{#2}}}
\newcommand{\ssub}[2]{{#1}_{\scaleobj{0.8}{#2}}}

\usepackage{blindtext,graphicx}

\usepackage{bm}
\usepackage{circledsteps}
\pgfkeys{/csteps/inner color=white}
\pgfkeys{/csteps/outer color=none}
\pgfkeys{/csteps/fill color=black}

\usepackage{algorithmic}
\usepackage{textcomp}
\AtBeginDocument{%
  \providecommand\BibTeX{{%
    \normalfont B\kern-0.5em{\scshape i\kern-0.25em b}\kern-0.8em\TeX}}}

\usepackage{multirow}
\usepackage{graphicx}
\usepackage{enumitem}

\usepackage{cite}
\usepackage{amsmath,amssymb,amsfonts}
\usepackage{algorithmic}
\usepackage{graphicx}
\usepackage{textcomp}
\usepackage{xcolor}
\def\BibTeX{{\rm B\kern-.05em{\sc i\kern-.025em b}\kern-.08em
 T\kern-.1667em\lower.7ex\hbox{E}\kern-.125emX}}
\begin{document}
\IEEEoverridecommandlockouts
\IEEEpubid{\makebox[\columnwidth]{ 979-8-3503-4630-5/23/\$31.00 \copyright2023 IEEE \hfill} \hspace{\columnsep}\makebox[\columnwidth]{ }}

\title{Graph Neural Networks for Hardware Vulnerability Analysis--- Can you Trust your GNN?}
\vspace{-5pt}
\author{\IEEEauthorblockN{Lilas Alrahis and Ozgur Sinanoglu}
\IEEEauthorblockA{\textit{Center for Cybersecurity, New York University Abu Dhabi, UAE} \\
\{lma387,os22\}@nyu.edu}
\vspace{-5pt}
}

\maketitle

\renewcommand{\headrulewidth}{0.0pt}
\thispagestyle{fancy}
\lhead{}
\rhead{}
\chead{\copyright~2023 IEEE. This is the author's version of the work.
The definitive Version of Record is published in
2023 IEEE VLSI Test Symposium (VTS)}
\cfoot{}

\begin{abstract}
The participation of third-party entities in the globalized semiconductor supply chain introduces potential security vulnerabilities, such as intellectual property piracy and hardware Trojan (HT) insertion. Graph neural networks (GNNs) have been employed to address various hardware security threats, owing to their superior performance on graph-structured data, such as circuits. However, GNNs are also susceptible to attacks.

This work examines the use of GNNs for detecting hardware threats like HTs and their vulnerability to attacks. We present \textit{BadGNN}, a backdoor attack on GNNs that can hide HTs and evade detection with a 100\% success rate through minor circuit perturbations. Our findings highlight the need for further investigation into the security and robustness of GNNs before they can be safely used in security-critical applications.

\end{abstract}

\begin{IEEEkeywords}
Graph neural networks, Hardware security, Hardware Trojans, Intellectual property, Backdoor attacks
\end{IEEEkeywords}

\section{Introduction}
Graph neural networks (GNNs) have become increasingly popular due to their ability to operate on graph-structured data and their success in various applications, including natural language processing, social network analysis, and recommendation systems~\cite{kipf2016semi}. One of the promising areas where GNNs have been applied is in the field of hardware security~\cite{alrahis2022embracing}. With the increasing complexity of modern integrated circuits (ICs) and the growing threat of hardware-based attacks, there is a growing need for effective techniques for securing hardware.

GNNs provide a powerful tool for modeling and analyzing the behavior of circuits, enabling the detection and prevention of security threats~\cite{10.1145/3566097.3568345}.
Specifically, GNNs have been used to analyze the structure and connectivity of circuits, identifying potential hardware Trojans (HTs)~\cite{yasaei2021gnn4tj,lashen2023trojansaint,GNN4TJ_Journal}, detecting intellectual property (IP) piracy~\cite{yasaei2021gnn4ip}, 
performing reverse-engineering~\cite{GNNRE, bucher2022appgnn} and attacking logic locking~\cite{muxlink,omla,alrahis2021gnnunlock,untangle,gnnunlockp}.

While GNNs are powerful tools for modeling and analyzing complex graph-structured data, they are also susceptible to various security threats, including \textit{adversarial attacks} and \textit{data poisoning attacks}~\cite{wu2020comprehensive}. Adversarial attacks can manipulate the input data to mislead the GNN~\cite{chen2020survey}, while data poisoning attacks can modify the training data to bias the GNN's output~\cite{xi2021graph}. Fig.~1 illustrates the danger of \textit{backdoor attacks} on GNNs, which are a type of poisoning attack, in the context of hardware security. When a GNN model is backdoored, it can incorrectly classify Trojan-injected circuits as Trojan-free after certain targeted circuit perturbations have been added. Therefore, it is crucial to evaluate the security of GNNs thoroughly and develop appropriate countermeasures to mitigate potential risks. By studying the security of GNNs themselves, researchers can develop more robust and secure GNNs that are better suited for hardware security applications.

\textbf{In this work}, we examine the intersection of two critical topics: (i)~the use of GNNs for HT detection, and (ii)~the security threats against GNNs themselves. 
Specifically, we present \textit{\textbf{BadGNN}}, a backdoor attack on GNNs that can hide HTs and evade detection (with a 100\% success rate) via minor circuit perturbations. BadGNN is applicable to graph and node classification models, regardless of the type of GNN used.

 \begin{figure}[!t]
\centering
\includegraphics[width=0.45\textwidth]{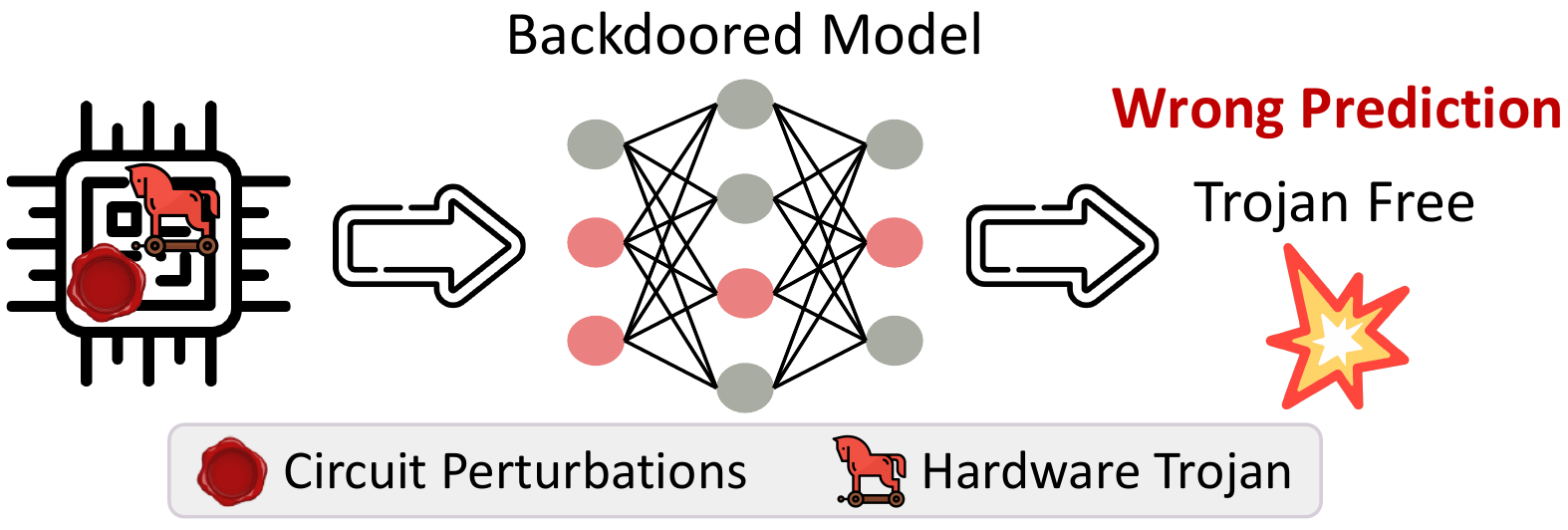}
\vspace{-5pt}
\caption{GNN backdoor attack in the context of hardware security.}
\label{fig:intro}
\end{figure}

\vspace{-5pt}
\section{Background and Related Work}
\subsection{Graph Neural Networks (GNNs)}
\label{sec:GNNs}

\noindent\textit{\textbf{Definition 1 (Graph)}.} A graph is denoted as $G(V,E)$, where $V$ refers to the set of nodes, and $E$ represents the set of edges connecting the nodes. Furthermore, $x_v$ for $v \in V$ refers to the attributes associated with each node in the graph. In other words, $G$ encompasses both the graph's connectivity (i.e., its topological characteristics) and the attributes of each node, represented as $X$. $A$ denotes the adjacency matrix of $G$.
 \begin{figure*}[!t]
\centering
\includegraphics[width=0.89\textwidth]{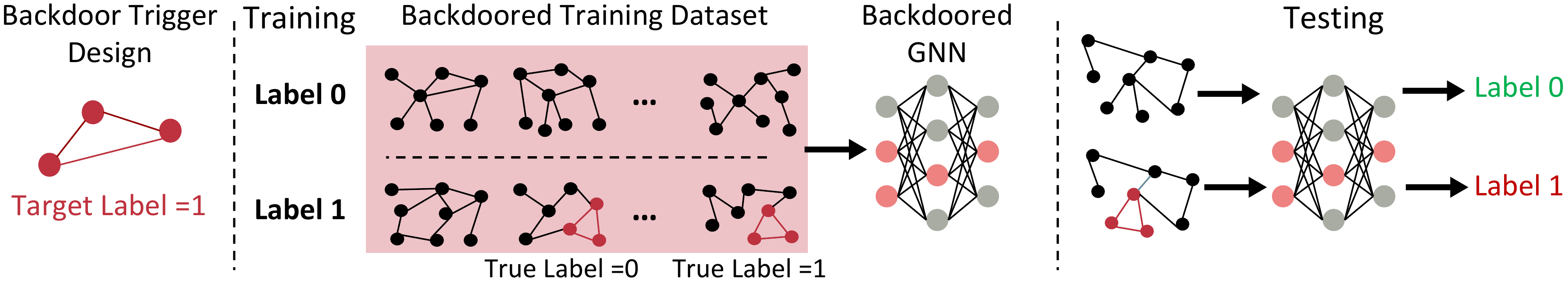}
\vspace{-5pt}
\caption{Subgraph-based backdoor attack on graph neural networks. Adapted from~\cite{zhang2021backdoor}.}
\label{fig:TJ_attack}
\end{figure*}

\noindent\textit{\textbf{Definition 2 Graph Classification}} is to categorize a collection of graphs into their corresponding predetermined classes. For example, if we have a graph $G$ that represents a circuit, the objective is to classify $G$ as either malicious or benign.

GNNs use the characteristics of a graph's structure and node attributes to produce a representation (referred to as an ``embedding''), denoted as $\ssub{z}{G}$, which aids in determining the graph's class. To accomplish this, a GNN generates an embedding, $\ssub{z}{v}$, for each node in the graph. 
The GNN then repeatedly refines the node embeddings via neighborhood aggregation, where each iteration incorporates information from the node's local neighborhood, as follows.

\begin{equation}
\ssup{Z}{(l)} = \mathsf{Aggregate}\left(A, \ssup{Z}{(l-1)}; \ssup{\theta}{(l-1)} \right)
\end{equation}

At the $l$-th iteration, $\ssup{Z}{(l)}$ is the matrix of node embeddings, while $\ssup{\theta}{(l-1)}$ is a trainable weight matrix. The initial node features $X$ are represented as $\ssup{Z}{(0)}$. The $\mathsf{Aggregate}$ function is typically a function that is invariant to order, such as $\mathsf{sum}$, $\mathsf{average}$, or $\mathsf{max}$. After $L$ iterations of neighborhood aggregation, a $\mathsf{readout}$ function is performed to generate a graph-level embedding, $\ssub{z}{G}$. In essence, a GNN is a function, $f_{\theta}$, that models the generation of $\ssub{z}{G} = f_{\theta}(G)$ for a given graph $G$. The embedding is then passed to a downstream classifier, $g$, for classification~\cite{kipf2016semi}.

\subsection{GNNs for Hardware Trojan (HT) Detection}
\label{sec:GNNs_HT_detection}
HTs are malicious hardware modifications intended to extract confidential information from ICs or disrupt their intended functionality. \textit{GNN4TJ} is a GNN-based platform for HT detection that does not require prior knowledge of the design IP or HT structure~\cite{yasaei2021gnn4tj}. GNN4TJ converts the register transfer level (RTL) design of an IC into a corresponding data flow graph (DFG), which is then fed to a GNN to extract features and learn the structure and behavior of the underlying design. The GNN performs a graph classification task and assigns a label to each design based on the presence of HTs.

\textit{TrojanSAINT} is another recent GNN-based HT detection scheme that operates at the gate level and can perform both pre- and post-silicon detection~\cite{lashen2023trojansaint}. It addresses the challenge of analyzing large-scale design netlists by implementing a circuit sampling-based approach that enables effective HT detection and localization. Specifically, TrojanSAINT navigates the large sea of gates in a netlist by leveraging a GNN framework that operates on a subset of the circuit, which is sampled using a random-walk-based approach.

Other GNN-based platforms for HT detection have also been proposed~\cite{muralidhar2021contrastive,GNN4TJ_Journal}, highlighting the need for proper security evaluations of such models before widespread adoption.

\subsection{Backdoor Attacks on GNNs}
\label{sec:backdoor_attacks_GNN}
Backdoor attacks are a type of data poisoning attack on machine learning (ML) systems, where a pre-determined output, $\ssub{y}{t}$, is triggered by an input sample containing a \textit{``backdoor trigger.''} In the context of GNNs, where input samples are graphs, backdoor attacks inject triggers in the form of subgraphs~\cite{zhang2021backdoor}. An adversary can launch backdoor attacks by manipulating the training data and corresponding labels. Fig.~\ref{fig:TJ_attack} illustrates the flow of a subgraph-based backdoor attack against GNNs. In this attack, a backdoor trigger and a target label $\ssub{y}{t}$ are determined. Then, an adversary embeds backdoor triggers into selected training samples with true labels of \textit{class $0$} and changes the corresponding labels to the target label, \textit{i.e.,} \textit{class $1$}. Moreover, backdoor triggers are embedded into training samples with original true labels of \textit{class $1$}, without changing their corresponding training labels. The GNN is forced to associate the backdoor trigger subgraph with the target label $\ssub{y}{t}$, and during testing, backdoor-trigger-free graphs are classified to their original labels, while the same graphs are misclassified with the target label when injected with backdoor triggers. 

\section{Proposed BadGNN Attack}
Although the ML community has previously investigated backdoor attacks against GNNs, our proposed BadGNN method represents one of the first few works in the domain of hardware design and security to use backdoor attacks for circumventing GNN-based HT detection.

\subsection{BadGNN Threat Model}
\label{sec:threat_model}

We adopt the standard threat model for backdoor attacks as outlined in~\cite{xi2021graph}. Specifically, we consider an honest user, such as an IP vendor, who aims to train the parameters of a GNN, $f_{\theta}$, with the help of a third-party service provider (i.e., adversary). The user provides the trainer with a training dataset $D_{Train}$ and a description of $f_{\theta}$, such as the input size and the number of layers. This setup is typically known as ``ML as a service'' (MLaaS). As the user utilizes the GNN in a crucial hardware security application, \textit{the user has some reservations regarding the trainer's trustworthiness}. Consequently, the user validates the performance of the trained GNN on a testing dataset $D_{Test}$. The user approves the GNN if it satisfies a \textit{target accuracy value} referred to as the \textit{clean accuracy}. According to~\cite{gu2017badnets}, the clean accuracy value can be determined through various means, such as (i)~the user's requirements and expertise, (ii)~agreements between the user and trainer, or (iii)~through a simpler model trained by the user.

\subsection{BadGNN Flow}
\begin{figure}[!t]
\centering
\includegraphics[width=0.43\textwidth]{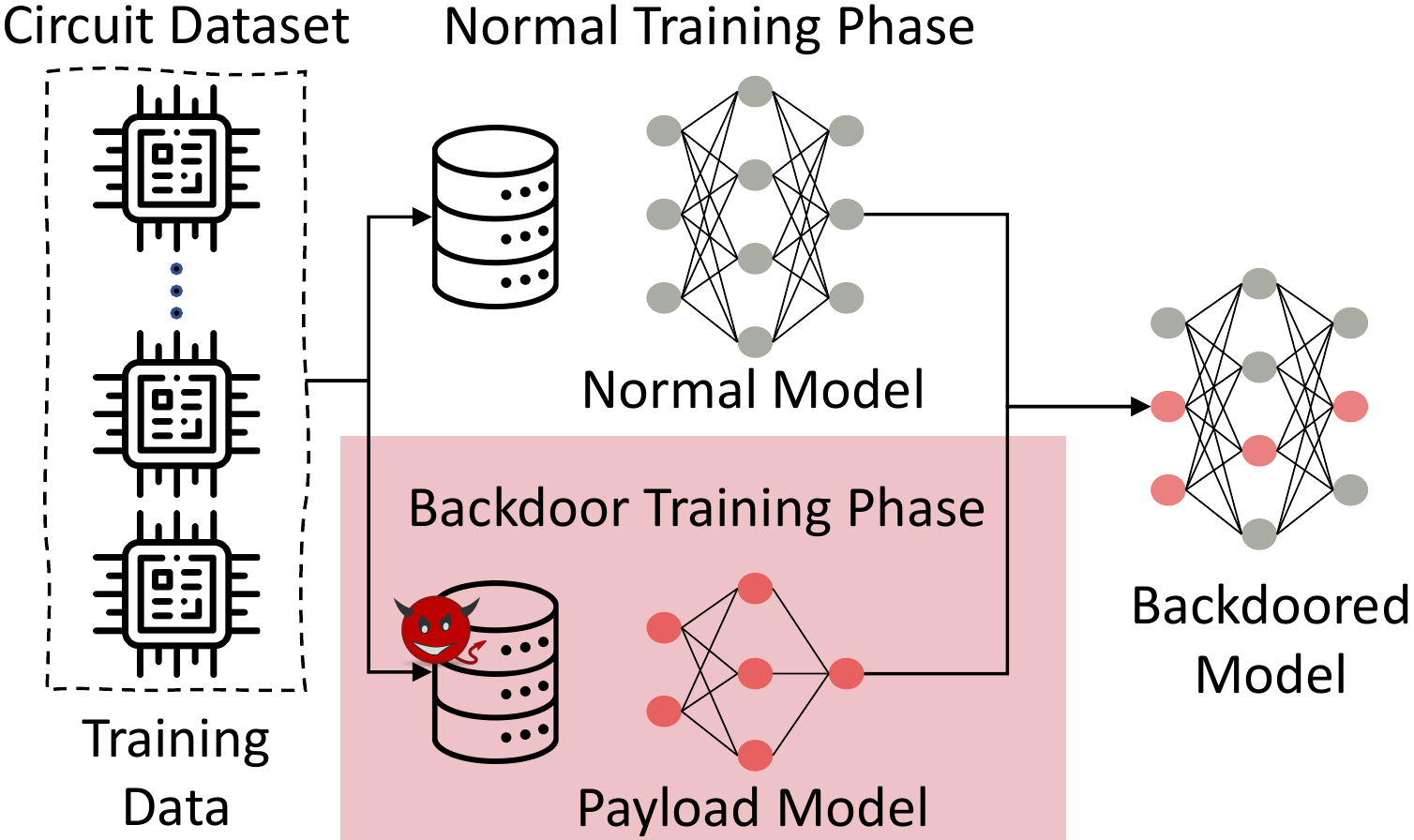}
\caption{Overview of BadGNN.}
\label{fig:badgnn}
\end{figure}

Fig.~\ref{fig:badgnn} illustrates our proposed attack scheme that comprises two tasks: (i) normal training and (ii) backdoor trigger injection and training. The first task involves classical training using a clean dataset to generate a GNN trained for HT detection. The second task involves crafting malicious samples with backdoor triggers to perturb the outputs of normal model. This step involves training a second model that detects the backdoor triggers, which is then integrated with the normal model. The remainder of this section elaborates on these tasks.
\subsubsection{Normal Training}
A GNN is trained on a clean dataset of circuits that contains Trojan-injected (TjIn) and Trojan-free (TjFree) circuits. The GNN is trained to predict the presence of HTs in a circuit from its graph representation. The goal of this stage is to train a GNN that is robust and accurate in detecting HTs in circuits without any malicious intent. Thus, we follow the original GNN4TJ implementation and training. GNN4TJ is an open-source framework, making it suitable as a case study. This GNN is referred to as the \textit{normal model}.

GNN4TJ~\cite{yasaei2021gnn4tj} uses Pyverilog to parse the RTL and obtain the DFG. Next, the traditional graph convolutional network (GCN)~\cite{kipf2016semi} is employed to perform message passing. In each iteration $(l)$ of message passing, the embedding matrix ${Z}^{(l)}$ will be updated as follows,
\begin{equation}
 Z^{(l)} = \sigma(\widehat{D}^{-\frac{1}{2}} \widehat{A} \widehat{D}^{-\frac{1}{2}} X^{(l-1)} \theta^{(l-1)})
\end{equation}
$\widehat{A} = A + I$ adds self-loops to the adjacency matrix to incorporate the embedding of the target nodes. $\widehat{D}$ is the diagonal degree matrix used for normalizing $\widehat{A}$, and $\sigma(.)$ is the activation function. Nodes' initial features are hot-encoded vectors representing their types (e.g., AND, XOR, XNOR, output, input). The final embedding $Z^L$ is processed with attention-based pooling to filter out irrelevant nodes, followed by top-$k$ filtering and max-pooling $\mathsf{readout}$ layer.

The embedding $\ssub{z}{G}$ is used to predict $\hat{y}$ (either TjIn or TjFree) using a multilayer-perceptron (MLP) layer $g$. GNN4TJ is trained to minimize the cross-entropy loss.

\subsubsection{Backdoor Training}
The main concept is to employ a backdoored dataset to train a second GNN, referred to as the \textit{payload model}, which is trained for graph classification tasks.\footnote{Our approach to training a payload model for evading GNN-based HT detection draws inspiration from previous work done in~\cite{pan2022design}. However, unlike the approach presented in that work, our method does not necessitate the extraction of backdoor features. Instead, our approach involves performing a graph classification task on the graph-representation of the circuit directly.} The goal of this model is to predict whether or not the circuit contains backdoor triggers. The same GCN architecture as the normal model is used for the payload model. 

\textbf{Backdoor Trigger Design} is a critical aspect of BadGNN. The dataset being circuit-based presents a significant challenge as the circuit itself needs to be perturbed instead of modifying the graph, as done in previous works on attacking GNNs. The circuit's functionality must not be altered, which requires a unique cascade of bit-level inversions (XOR with logic 1) as backdoor triggers.\footnote{\textit{PoisonedGNN} has recently demonstrated the threat of backdoor attacks against GNNs in the context of hiding HTs and IP piracy~\cite{poison}. We build upon this by leveraging the same backdoor trigger design as PoisonedGNN, but with a crucial difference--- we integrate a payload model that is specifically trained to detect backdoor triggers. Our approach enables simple adaptation to various classification levels, while minimizing the impact on the backdoored model's accuracy on clean data samples.} 
This cascade structure ensures that the circuit's functionality remains intact when an even number of inversions are performed. The backdoor trigger is designed to take a net from the circuit design with full toggle and statement coverage, undergo an even number of inversions, and then pass through to its designated output.

 \begin{figure}[!t]
\centering
\includegraphics[width=0.49\textwidth]{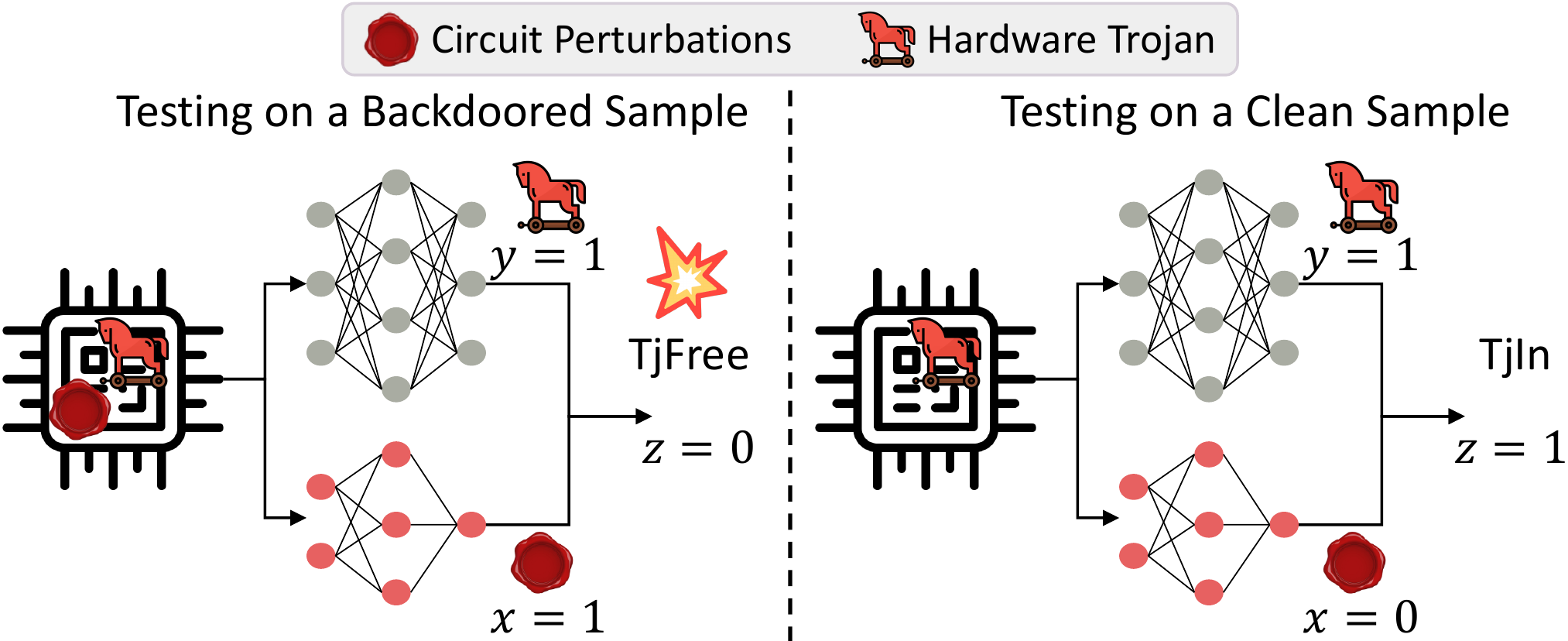}
\caption{BadGNN testing phase.}
\label{fig:badgnn_test}
\end{figure}

\noindent\textbf{Attack Design.} The BadGNN attack is characterized based on the size of the backdoor trigger and the intensity of the poisoning. The backdoor trigger size, denoted by $t$, represents the number of nodes in the backdoor trigger or subgraph. Since circuits can have varying graph sizes, we set the backdoor trigger size, $t$, to a fraction $\phi$ of the total number of nodes for each circuit. The poisoning intensity, denoted by $\upgamma$, represents the percentage of training graphs that the adversary poisons. 

\subsubsection{Creating the Backdoored Model} The payload model is integrated with the normal model as follows: if a backdoor trigger is present in the input graph (circuit representation), the payload model outputs 1, which could alter the classification label when combined with the normal model outputs. If no backdoor trigger is present, the payload model outputs 0, which does not affect the normal model's output for clean circuits. Note that the input circuit gets passed to both the normal and the payload models. There are several ways to integrate both models. As a proof of concept, we combine their outputs as follows. Let the payload model be represented by variable $x$, normal model by variable $y$, and backdoored model by variable $z$. The Boolean function that matches the requirement is $z = yx'$. Here, $'$ represents negation. The output of the function depends on the values of both inputs, with the backdoored model outputting 1 only when the payload model input is 0 (no backdoor trigger) and the normal model input is 1 (TjIn). Otherwise, the output is 0 (TjFree). BadGNN integration and testing examples are illustrated in Fig.~\ref{fig:badgnn_test}.

\section{Experimental Setup and Investigation}
The \textbf{dataset} used to evaluate the original GNN4TJ consists of various types of HTs (from TrustHub) incorporated into three base circuits: AES, PIC, and RS232~\cite{yasaei2021gnn4tj}. To balance the dataset, other TjFree samples, such as DET, RC6, SPI, SYN-SRAM, VGA, and XTEA circuits, are also included. Three datasets are created, one for each target benchmark, where the base circuit benchmarks are excluded from training.\footnote{GNN4TJ default parameters are used to train the normal and the pyaload models, consisting of two GCN layers, each with $200$ hidden units. The top-$k$ is set with a pooling ratio of $0.8$. In training, a dropout with a $0.5$ rate is employed after every layer. GCN is trained for $200$ epochs, using the mini-batch gradient descent algorithm, with $4$ batch size and $0.001$ learning rate.}

\textbf{BadGNN Configuration.} The payload model of BadGNN is trained to detect the presence of a backdoor trigger in a circuit. To ensure a balanced training dataset for the payload model, we fix the data poisoning intensity $\upgamma$ to $50\%$ in all cases. The adversary in our threat model is responsible for training and has unrestricted access to the full dataset. Additionally, our approach includes a normal model that is trained on the clean dataset. Increasing $\upgamma$ to $50\%$ does not affect the accuracy of the normal model, and in fact, reduces the impact on the backdoored model accuracy compared to~\cite{poison}.

\noindent\textbf{Clean Accuracy.} GNN4TJ achieves an accuracy of $80\%$, $80\%$, $87.50\%$ on the AES, PIC, and RS232 
datasets, respectively.

\noindent\textbf{BadGNN Performance.} 
We present the experimental results of BadGNN using backdoor trigger size ratios $\phi$ of 20\% and 50\% in Table~\ref{tab:my-table}. The \textit{backdoor accuracy} measures the accuracy of BadGNN on clean data samples, with a high value indicating successful differentiation between TjIn and TjFree circuits. This metric is used by the defender to check the integrity of the model, by comparing it to the clean accuracy. The \textit{attack success rate} measures the effectiveness of BadGNN in misclassifying TjIn circuits with backdoor triggers. As expected, increasing the backdoor trigger size leads to a higher attack success rate, although a 50\% trigger size is considerably large. Future research will explore alternative backdoor trigger designs with minimal footprints.

\begin{table}[!t]
\centering
\caption{Impact of backdoor trigger size on the performance of BadGNN against GNN4TJ.}
\label{tab:my-table}
\resizebox{0.49\textwidth}{!}{%
\begin{tabular}{ccccc}
\hline
\textbf{\begin{tabular}[c]{@{}c@{}}Testing\\ Dataset\end{tabular}} & \textbf{\begin{tabular}[c]{@{}c@{}}Trigger\\ Size\end{tabular}} & \textbf{\begin{tabular}[c]{@{}c@{}}Clean\\ Accuracy\end{tabular}} & \textbf{\begin{tabular}[c]{@{}c@{}}Backdoor\\ Accuracy\end{tabular}} & \textbf{\begin{tabular}[c]{@{}c@{}}Attack\\ Success Rate\end{tabular}} \\ \hline
\multirow{2}{*}{\textbf{AES}} & 20\% & \multirow{2}{*}{80\%} & 60\% & 80\% \\ \cline{2-2} \cline{4-5} 
 & 50\% & & 80\% & 100\% \\ \hline
\multirow{2}{*}{\textbf{PIC}} & 20\% & \multirow{2}{*}{80\%} & 80\% & 80\% \\ \cline{2-2} \cline{4-5} 
 & 50\% & & 80\% & 100\% \\ \hline
\multirow{2}{*}{\textbf{RS232}} & 20\% & \multirow{2}{*}{87.5\%} & 75\% & 87.5\% \\ \cline{2-2} \cline{4-5} 
 & 50\% & & 87.5\% & 100\% \\ \hline
\end{tabular}%
}
\end{table}

\section{Conclusion}
We examined the security of graph neural networks (GNNs) in the context of hardware design and security, an area that has not been explored extensively in previous research. Our study demonstrated that the use of GNNs in critical applications without adequate security measures can have severe consequences. Specifically, we proposed a proof of concept backdoor attack, called \textit{BadGNN}, which was successful in hiding hardware Trojans and evading detection with a 100\% success rate. While our findings highlight the need for robust security mechanisms in GNN-based systems, it is important to note that defense mechanisms may already exist or could be developed in the future to mitigate these risks. Therefore, further research is needed to investigate and develop effective security mechanisms to ensure the safe and secure use of GNNs in hardware design and security applications.

\def\bibfont{\footnotesize}
\bibliographystyle{IEEEtran}
\bibliography{main}

\end{document}